\documentclass[% 
reprint,
superscriptaddress,
 amsmath,amssymb,
 aps,
 pre,
floatfix,
]{revtex4-2}
\usepackage{graphicx}% Include figure files
\usepackage{dcolumn}% Align table columns on decimal point
\usepackage{bm}% bold math
\usepackage[bordercolor=white,backgroundcolor=gray!30,linecolor=black,colorinlistoftodos]{todonotes}

\usepackage{subfigure}
\usepackage[figuresright]{rotating}
\usepackage{float}
\begin{document}

%\preprint{APS/123-QED}

\title{Effect of the degree of an initial mutant in Moran processes in structured populations}
% Force line breaks with \\
%\thanks{A footnote to the article title}%
\author{Javad Mohamadichamgavi}\thanks{jmohamadi@mimuw.edu.pl}
%\affiliation{Institute of Applied Mathematics and Mechanics, University of Warsaw, ul. Banacha, 2 02-097 Warsaw, Poland}
\author{Jacek Mi\c{e}kisz}\thanks{miekisz@mimuw.edu.pl}
\affiliation{Institute of Applied Mathematics and Mechanics, University of Warsaw, ul. Banacha 2, 02-097 Warsaw, Poland}

%\date{\today}

\begin{abstract}
We study effects of the mutant's degree on the fixation probability, extinction, and fixation times, in Moran processes on Erd\"{o}s-R\'{e}nyi and Barab\'{a}si-Albert
graphs. We performed stochastic simulations and used mean-field type approximations to obtain analytical formulas. We showed that the initial placement of a mutant
has a significant impact on the fixation probability and extinction time, while it has no effect on the fixation time. In both types of graphs, 
an increase in the degree of an initial mutant results in a decreased fixation probability and a shorter time to extinction. Our results extend previous ones to arbitrary
fitness values.
\end{abstract}

%\begin{description}
%\item[PACS numbers] 87.23.Kg
%\end{description}

%\pacs{Valid PACS appear here}
\maketitle
%\tableofcontents

% *********************************************************************
%\section{\label{intro}Introduction}
\section{Introduction}

Extinction and fixation of phenotypes and behaviors in animal and human populations are important features of evolutionary processes. 
The classical model here is the Moran process in populations of fixed finite sizes, consisting of individuals equipped with one of two phenotypes \cite{moran}.
Namely, an individual is chosen proportional to its fitness and it gives birth to an offspring who inherits its phenotype and replaces a randomly chosen individual. 
Such a birth-death process (also called an invasion process) is a Markov chain with two absorbing states: homogeneous populations with only one phenotype present. 
One may also consider death-birth Moran process, where an individual is chosen with a probability inversely proportional to its fitness and is replaced by a randomly
chosen individual \cite{Antal,nowakbook1,Pavlogiannis,Traulsen22}.

Here we will study fixation probability and expected fixation and extinction times of mutants in birth-death Moran processes on graphs.
In such processes, offspring replace neighbors of ancestors. The study of fixation probabilities on graphs was initiated by Lieberman, Hauert, and Nowak in \cite{lieberman}. 
Since then there appeared many papers on Moran processes on graphs
\cite{ohtsuki,traulsen2,Antal2,Sood,broom2,altrock,mark,Shakarian,allen,traulsen3,lu,Allen3,maciejewski,askari2,Adlam,Hindersin,Cuesta2,askari,Cuesta,mehdi,strogatz,moler,Tkadlec,Allen4,Allen2,McAvoy,dehgha,yagoobi,mehdi2}. 
They have shown that the structure of the population can have a significant impact on the evolutionary dynamics 
and can affect fixation probability and fixation time. 

In particular, Antal, Redner, and Sood \cite{Antal2} studied fixation probability on degree-heterogeneous power-law graphs. 
They considered various dynamics and derived approximate analytical formulas for the dependence of the fixation probability 
on the degree of an initial mutant. In particular, for the birth-death process (biased invasion process) 
in the case where fitness of a mutant is very close to that of a resident, they showed that the fixation probability of a mutant is inversely proportional to its degree. 
In the follow-up paper \cite{Sood}, they studied fixation time in the neutral case. For similar results see also \cite{maciejewski}.

The effect of the placement of an initial mutant was also studied by Broom, Rychtář, and Stadler \cite{mark}.  
The authors derived a useful analytical approximation of fixation probability as a function of the degree of the initial mutant for some small graphs. 
Their findings revealed that initial mutants with fewer connections had a greater advantage, resulting in a higher probability of fixation. 

Here we present results of stochastic simulations and use a mean-field type approximation to derive analytical formulas 
for the fixation probability on Erd\"{o}s-R\'{e}nyi \cite{er} and Barab\'{a}si-Albert \cite{ba1,ba2}
graphs which agree with stochastic simulations for any fitness values. We also derive formulas for fixation and extinction time for arbitrary fitness values.

We have extended results of \cite{Antal2,Sood,maciejewski} to models with arbitrary fitness values. In particular, we compared our approximate formula
for the fixation probability to that in \cite{Antal2}, see Appendix B. In general, we showed that the initial placement of a mutant has a significant impact 
on the fixation probability and extinction time, while the degree of the initial mutant has no effect on the fixation time. 
In both types of graphs, an increase of the degree of the initial mutant results in a decreased probability of the mutant taking 
over the entire population and a shorter time to extinction. 

In the two graphs with the same size and average vertex degree, for an initial mutant with the same degree, 
the fixation probability in the Erd\"{o}s-R\'{e}nyi graph is higher than in the Barab\'{a}si-Albert one. 
However, due to the presence of hubs, the average fixation probability in the Barab\'{a}si-Albert graph is higher.
 
In Section II, we present our model - a Markov chain of the Moran process on graphs. In Section III, we present results of stochastic simulations
and derive analytical results. Further analysis of extinction time is contained in Section IV. Discussion follows in Section V.

\section{Models and methods}
\label{start}

We will study here the Moran process on two random graphs, the Erd\"{o}s-R\'{e}nyi (ER) \cite{er} 
and the scale-free Barab\'{a}si-Albert (BA) one \cite{ba1,ba2}. We build the ER graph by putting 
with probability $p$ an edge between every pair of $N=500$ vertices. 
It follows that the average degree of vertices (the average number of neighbors) is equal to $\alpha=p(N-1).$ 
The BA network is built by the preferential attachment procedure. 
We start with $m_{o}$ fully connected vertices and then we add $N-m_{o}$ vertices, each time connecting them with $m$ already available vertices 
with probabilities proportional to their degrees. If $m_{o}=\alpha+1$ and $m=\alpha/2$, then we get a graph with the average degree equal to $\alpha$. 
It is known that such a graph is scale-free with the probability distribution of degrees given by $p(k)\sim k^{-3}$ \cite{ba1,ba2,durett}.

Individuals - mutants with a constant fitness $r$ and residents with a constant fitness $1$ - are located on the vertices of a graph. 
At every time step of the process, an individual is chosen with a probability proportional to its fitness and then gives birth to an offspring 
who inherits its phenotype and replaces a randomly chosen individual. Such a Markov chain with ${2}^{N}$ states has two absorbing ones: 
homogeneous populations with only one type of individuals present.
We start our process with one mutant in the population of residents. We are interested in fixation probability, 
expected time to fixation and extinction of mutants, denoted respectively by $\rho^f_1, e^f_i$ , and $e^e_1$. 
To find such quantities, we performed stochastic simulations. 

We simulate the Moran process on graphs with $N=500$ vertices. For every $k$, we randomly place an initial mutant on one of the nodes of the degree k
and perform 10,000 Monte-Carlo simulations until the population reaches one of two absorbing states. To get fixation probability, we divide 
the number of times the process ends in the all-mutant state by 10,000. Finally, we calculate an average fixation (extinction) time for some chosen k's. 

It is impossible to get rigorous analytical expressions for fixation and extinction probabilities and times for Moran processes on graphs.
The situation is much simpler in well-mixed populations \cite{Diaz,Nie}.
Then the Moran process becomes a Markov chain with $N+1$ states corresponding to the number of mutants or more precisely, 
a random walk on $\{0, ... , N\}$
with state-dependent probabilities of moving to the right, $p_{i \rightarrow i+1}$, to the left, $p_{i \rightarrow i-1}$, and not moving,
$1 - p_{i \rightarrow i+1}-p_{i \rightarrow i-1}$. 
Such a model is well-known in the context of the gambler's ruin problem. 
It is easy to see that here we have that $p_{i\rightarrow i+1} = r p_{i\rightarrow i-1}$ where $r$ is the fitness of the mutant.
For the fixation probability starting from $i$ mutants, one gets that 

\begin{equation}\label{mixed}
\rho^f_i=\frac{1-\frac{1}{r^i}}{1-\frac{1}{r^N}}.
\end{equation}
\begin{figure*}
{\includegraphics[scale=0.29]{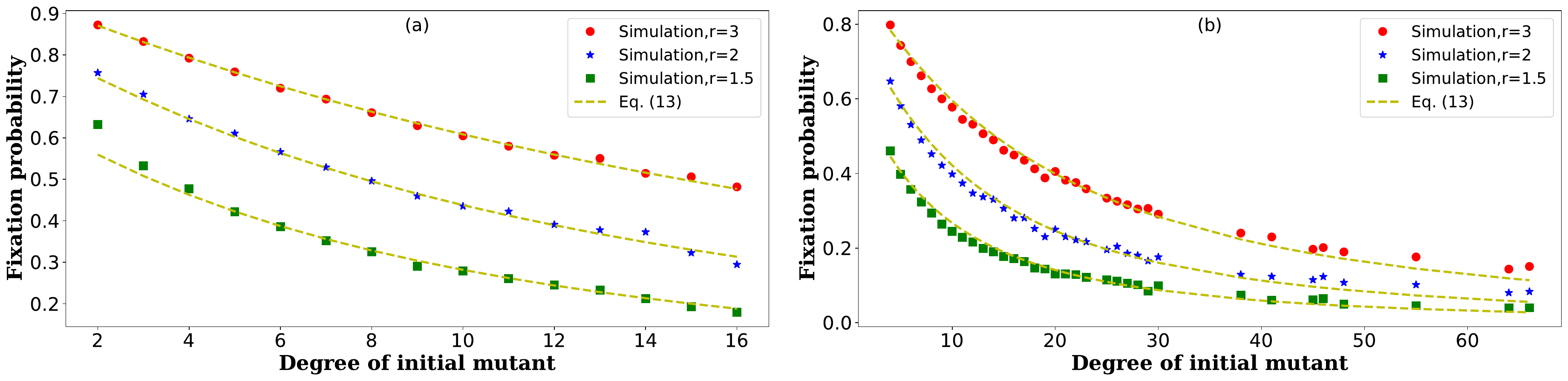}}
\centering
\caption{Fixation probability as a function of a degree of an initial mutant for three fitness values: 3, 2, 1.5 in (a) ER and (b) BA graphs with the size of 500 
and the average degree of 8.}
\label{fp1}
\end{figure*}
For the fixation probability on graphs, to take into account a degree of an initial mutant, we derive some approximate expressions for transition probabilities 
up to $3$ mutants and then use the elementary conditional probability formula and the fixation probability $\rho^f_3$ from (\ref{mixed})
to obtain an analytic expression for the fixation probability as a function of a degree of the initial mutant. 
In that way we passed from the full Markov chain with $2^N$ states to the one with $N+1$ states corresponding to the number of mutants.

In the following section, we derive analytic expressions for the fixation probability and expected fixation and extinction times, and compare them 
with results of stochastic simulations.
 
\section{Results}
\label{start}

\subsection{Fixation probability}

\begin{figure}
{\includegraphics[scale=0.50]{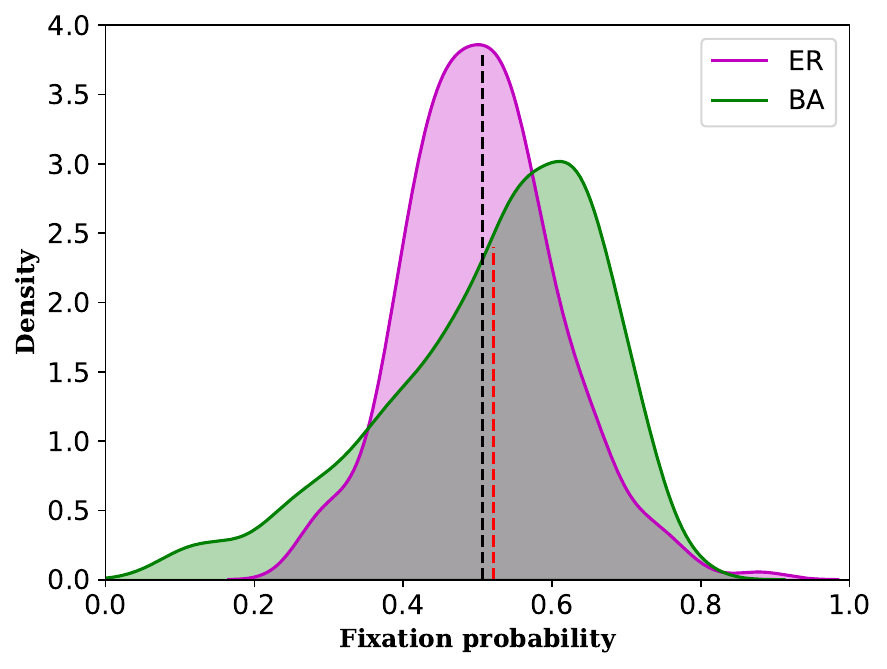}}
\centering
\caption{Distribution of fixation probabilities of a mutant with respect to initial placements on ER and BA graphs with the size of $500$ 
and the average degree of $8$. The fitness is $2$. Dash lines show the average fixation probability.}
\label{fp2}
\end{figure}
\begin{figure}
{\includegraphics[scale=0.50]{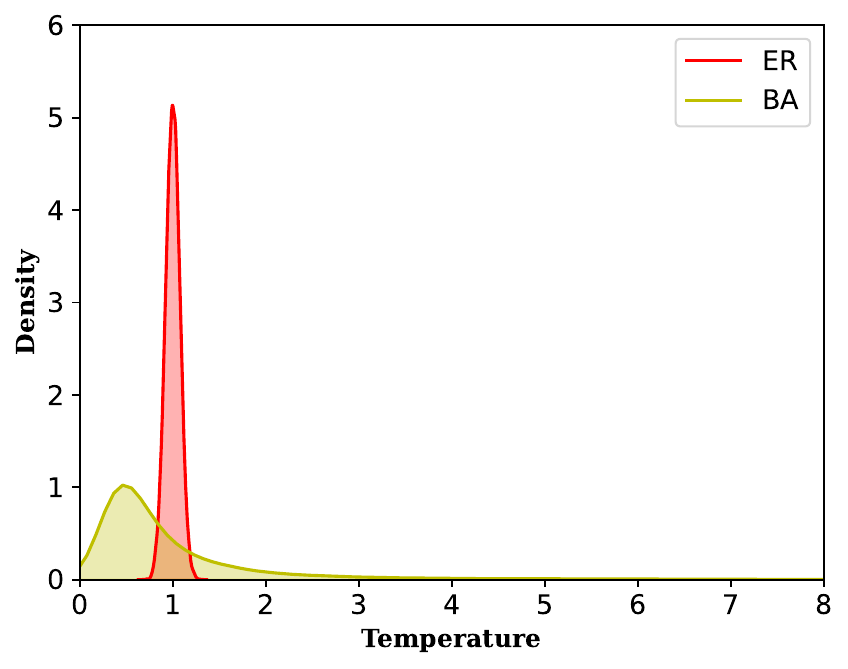}}
\centering
\caption{The distribution of temperatures in ER and BA graphs. The node temperature is simply defined as the inverse of the degree of all nodes leading to that node.}
\label{temp}
\end{figure}

Let $\rho^f_i$ be the fixation probability when initially there are $i$ mutants in the population. We denote by $P_{i \rightarrow j}$ 
the probability of the transition from the state $i$ to the state $j$; of course, for our Moran process, transition probabilities are non-zero only for $|i-j| \leq 1$. 
From the total probability formula we get

\begin{equation}
\rho^f_1=P_{1\rightarrow 1}\rho^f_1+P_{1\rightarrow 2}\rho^f_2,
\label{rho1}
\end{equation}

\begin{equation}
\rho^f_2=P_{2\rightarrow 1}\rho^f_1+P_{2\rightarrow 2}\rho^f_2+P_{2\rightarrow 3}\rho^f_3.
\label{rho2}
\end{equation}

We substitute (\ref{rho2}) to (\ref{rho1}) and get

\begin{equation}
\rho^f_1=\dfrac{P_{1\rightarrow 2}P_{2\rightarrow 3}}{(P_{1\rightarrow 2}+P_{1\rightarrow 0})(P_{2\rightarrow 1}+P_{2\rightarrow 3})-P_{1\rightarrow 2}P_{2\rightarrow 1}}\rho^f_3
\label{rho1n}
\end{equation}

For the number of mutants to increase from $1$ to $2$, the mutant should be chosen for reproduction, hence
\begin{equation}
P_{1 \rightarrow 2}=\frac{r}{N+r-1}.
\label{p12}
\end{equation}

For the number of mutants to decrease from $1$ to $0$, a resident from the mutant's neighborhood with degree $k$ must be selected 
and its offspring placed in the mutant's location so

\begin{equation}
P_{1 \rightarrow 0}=\sum_{k^{'}}^{}\frac{k}{r+N-1}\frac{1}{{k^{'}}} {{\rho}^{'}(k^{'})},
\label{tr10}
\end{equation}

where $k'$ is the degree of the selected resident and ${{\rho}^{'}(k^{'})}$ is the degree distribution of neighbors of the mutant.  

The distribution ${{\rho}^{'}(k^{'})}$ of degrees of a node which is connected to a given node is given by $\frac{k^{'}\rho(k^{'})}{<k>}$,
where $\rho(k^{'})$ is the degree distribution in the graph \cite{feld,newman}. Therefore 

\begin{equation}
P_{1 \rightarrow 0}=\frac{k}{r+N-1}\frac{1}{<k>}.
\label{tr10n}
\end{equation}

Now we have to calculate $P_{2\rightarrow 1}$. For this transition, one of the resident neighbors of two mutants must be chosen for the reproduction 
and replace the mutant. 
One of them is the first mutant with a degree of $k$, while the other is a previously mutated node with a degree of $k^{'}$. We have the following formula,

\begin{equation}
\begin{split}
 P_{2 \rightarrow 1}=\sum_{k^{'}}^{}\frac{k-1}{2r+N-2}\frac{1}{{k^{'}}} \frac{k^{'}\rho(k^{'})}{<k>}+\hspace{40pt} \\ \sum_{k^{'}}^{}\sum_{k^{"}}^{}\frac{k'-1}{2r+N-2}\frac{1}{{k^{"}}} \frac{k^{'}\rho(k^{'})}{<k>}\frac{k^{"}\rho(k^{"})}{<k>}.
\end{split}
\label{tr21}
\end{equation}

In this equation, the second moment $\sum_{k'}{k'}^2\rho(k')$ appears. After some simplifications we get

\begin{equation}
P_{2 \rightarrow 1}=\dfrac{1}{2r+N-2}\dfrac{1}{<k>}(k+\frac{<k^2>}{<k>}-2).
\label{tr21n}
\end{equation}

In the the same way we obtain $P_{2 \rightarrow 3}$, 
\begin{equation}
  P_{2 \rightarrow 3}=\frac{r}{2r+N-2}\frac{k-1}{k}+ \sum_{k^{'}}^{}\frac{r}{2r+N-2}\frac{k^{'}-1}{k^{'}} \frac{k^{'}\rho(k^{'})}{<k>}.
\label{tr23}
\end{equation}

and after some simplifications we have

\begin{equation}
P_{2 \rightarrow 3}=\dfrac{r}{2r+N-2}(\frac{k-1}{k}-\frac{1}{<k>}+1), 
\label{tr23n}
\end{equation}

This allow us to rewrite (\ref{rho1n}) as follows:
\begin{widetext}
\begin{equation}
\rho^f_1= \dfrac{{<k>}^2r^2(2-\frac{1}{k}-\frac{1}{<k>})}{(r<k>+k)(r<k>(2-\frac{1}{k}-\frac{1}{<k>})+(k-2+\frac{<k^2>}{<k>}))-r<k>(k-2+\frac{<k^2>}{<k>})}\rho^f_3.
\label{rho1na}
\end{equation}
\end{widetext}

Now we assume that the effect of an initial mutant is limited to transitions from one-mutant population to two-mutant population 
and also from two mutants to three mutants. Hence we take $\rho^f_3$ from (\ref{mixed}) and get

\begin{widetext}
\begin{equation}
\rho^f_1= \dfrac{{<k>}^2r^2(2-\frac{1}{k}-\frac{1}{<k>})}{(r<k>+k)(r<k>(2-\frac{1}{k}-\frac{1}{<k>})+(k-2+\frac{<k^2>}{<k>}))-r<k>(k-2+\frac{<k^2>}{<k>})}\dfrac{1-\frac{1}{r^3}}{1-\frac{1}{r^N}}.
\label{rho1final}
\end{equation}
\end{widetext}

To determine the fixation probability of mutants with degree $k$, we use (\ref{rho1final}), which requires the knowledge of the average degree $<k>$ 
and the second moment $<k^2>$ of the graph. For the Erd\"{o}s-R\'{e}nyi graph with the binomial degree distribution, 
the average degree is equal to $<k>=(N-1)p$, and the second moment  $<k^2>=<k>(<k>(1-1/(N-1))+1)$. For the Barab\'{a}si-Albert graph,
 the degree distribution follows a power-law, 
resulting in an average degree of $<k>=2m$ and $<k^2> \equiv <k>^2log N/4$ \cite{Alodjants}.

\begin{figure}
{\includegraphics[scale=0.5]{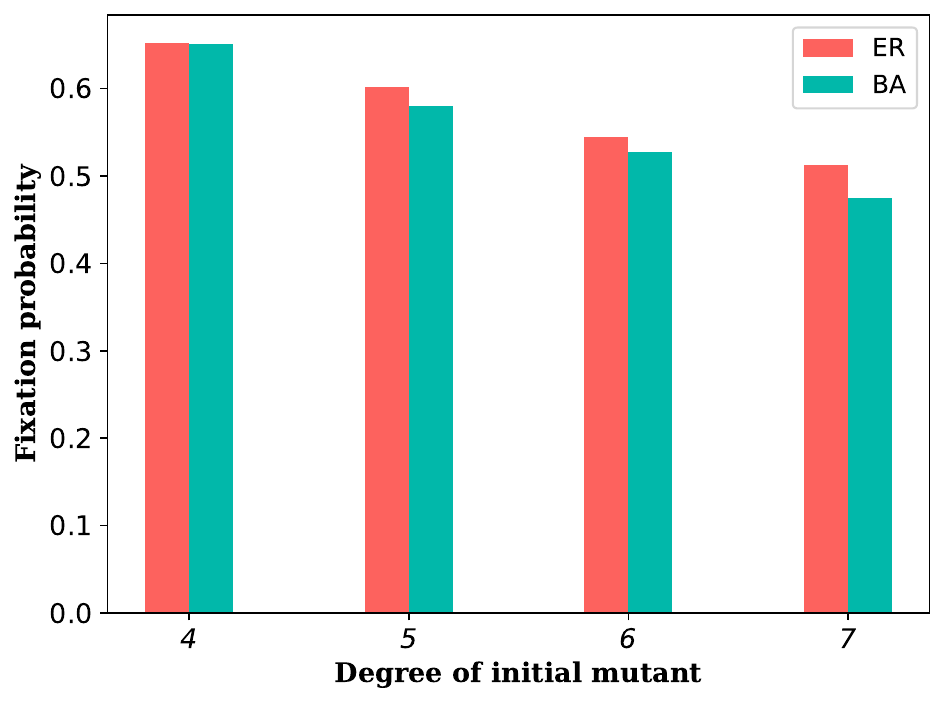}}
\centering
\caption{Fixation probability for various initial mutants with the same degree in ER and BA graphs. 
The size of the graph is $500$ with the average degree of $8$ and $r=2$.}
\label{fps}
\end{figure}
\begin{figure*}
{\includegraphics[scale=0.29]{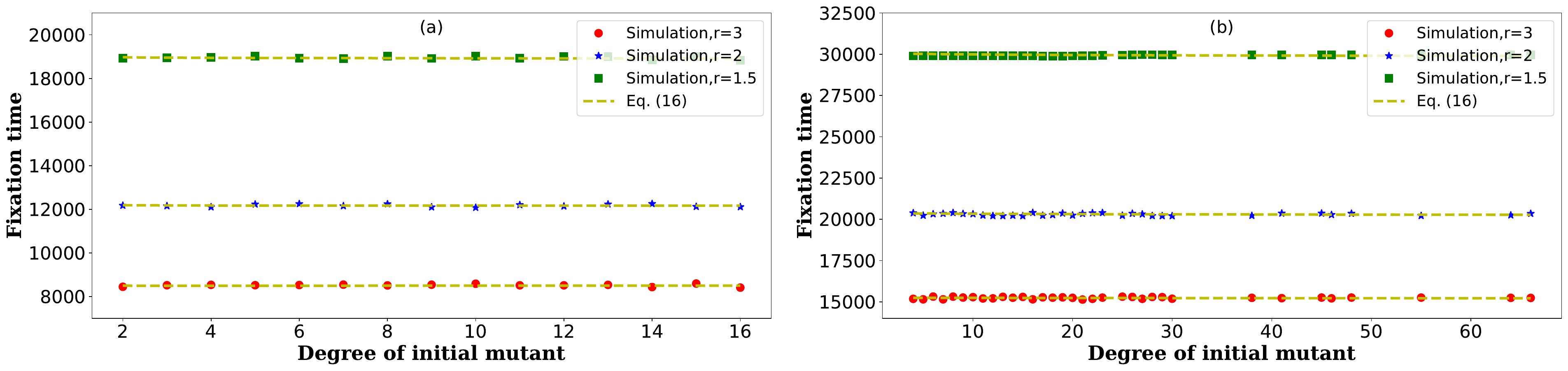}}
\centering
\caption{Fixation time as a function of the degree of an initial mutant for three fitness values, 3, 2, 1.5 in (a) ER and (b) BA graphs with the size of 500 
and the average degree of 8.}
\label{ft}
\end{figure*}
\begin{figure}
{\includegraphics[scale=0.50]{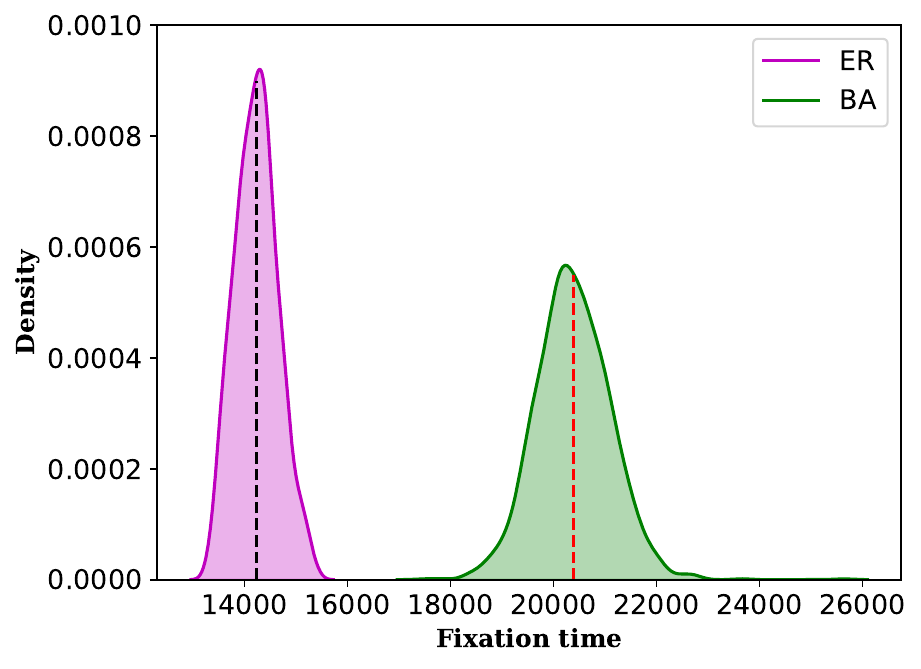}}
\centering
\caption{Distribution of fixation times of a mutant with respect to initial placements on ER and BA graphs with the size of $500$ 
and the average degree of $8$. The fitness is $2$. Dash lines show the average fixation time.}
\label{ft3}
\end{figure}
Fig. \ref{fp1} presents the results of stochastic simulations and our analytical approximation for the fixation probability. 
Our results show a decrease in fixation probability as the initial mutant's degree increases, which agrees with previous findings by Antal, Redner, and Sood \cite{Antal2}. 
They demonstrated that if fitness of a mutant is close to that of a resident, then the fixation probability is inversely proportional to the degree of the initial mutant. 
However, comparing our approximation to theirs reveals that their model accurately calculates fixation probability for mutants with fitness close to residents,
while our model maintains accuracy across all fitness levels, see Appendix B.

As the degree of the initial mutant increases, it becomes more probable that one of its resident neighbors replaces it.
Hence, vertices with a lower number of connections have a higher chance of remaining and progressing toward fixation. In the BA graph, 
the degree distribution follows a power law, which means that there are a few nodes with very high degrees and many nodes with low degrees.  
Consequently, in the BA graph, there may be initial nodes with a high probability of fixation, mainly those with low degrees,
and others with very low fixation probabilities, mainly those with high degrees, in contrast to the ER graph where the probability of fixation 
is more uniform across nodes.

In Fig. \ref{fp2}, we present the distribution of fixation probabilities for a mutant with respect to all initial placements in both graphs of the same size, 
average degree, and fitness. In ER graphs, fixation probabilities for all vertices are distributed approximately symmetrically 
around the average fixation probability. 
In BA graphs, the distribution is asymmetric. Moreover, we see that for $r=2$, the average fixation probability in the ER graph is approximately $0.5$ 
which is the fixation probability in a well-mixed population, while this value is slightly higher in the BA graph. 

This can be understood on the basis of a concept of temperature vertex which is defined as the sum of the inverses of degrees 
of all neighbors of a given vertex. It was proven in \cite{lieberman} that for graphs for which all vertices have the same temperature, 
the so-called isothermal graphs, the average fixation probability is equal to the one in well-mixed populations.

Fig. \ref{temp} displays the temperature distribution for both graphs with the size of 500 and an average degree of 8. 
The ER graph exhibits a peak at 1, indicating that nearly all nodes have the same temperature. In contrast to the ER graph, 
the BA graph's vertices show the large spectrum of temperatures. While there are many vertices with small degrees, leading to higher fixation probabilities, 
there are only a few vertices with large degrees, leading to low fixation probabilities. Consequently, the average fixation probability in BA graphs is higher 
than in well-mixed populations and ER graphs. 

In Fig. \ref{fps}, we present the fixation probability for an equal degree of initial placement for both graphs. 
As shown, the fixation probability of one mutant with the the same degree in both graphs, is slightly higher for the ER graph. 
However, this does not contradict the observation that the average fixation probability is higher in the BA network. 
While it is true that the fixation probability for initial mutants with the same degree is higher in ER graphs, there are more nodes with a low degree in BA graphs, 
which have a higher fixation probability than in ER graphs. As a result, the average fixation probability in BA graphs is greater than that in ER graphs.

\subsection{Fixation Time}

To derive an analytical expression for the expected value of the fixation time we proceed in an analogous way as for the fixation probability - 
we use the conditional expected value formula. We follow Antal and Sheuring \cite{Antal} and obtain an equation for $t_i^f$, 
fixation time in populations with $i$ mutants,

\begin{equation}
    \rho_1^f t_1^f=P_{1 \rightarrow 1}\rho_1^f (t_1^f+1)+P_{1 \rightarrow 2}\rho_2^f (t_2^f+1) 
    \label{fie1}
\end{equation}
and
\begin{equation}
    \rho_2^f t_2^f=P_{2 \rightarrow 1}\rho_1^f (t_1^f+1)+P_{2 \rightarrow 2}\rho_2^f (t_2^f+1)+P_{2 \rightarrow 3}\rho_3^f (t_3^f+1)
    \label{fie2}
\end{equation}

We have $\rho_2^f t_2^f$  from (\ref{fie2}), we put it to (\ref{fie1}) and after some simplifications we get

\begin{widetext}
\begin{align}
\begin{split}
   t_1^f&=\dfrac{P_{2 \rightarrow 1}+P_{2 \rightarrow 3}}{\rho_1^f((P_{1 \rightarrow 0}+P_{1 \rightarrow 2})(P_{2 \rightarrow 1}+P_{2 \rightarrow 3})-P_{2 \rightarrow 1}P_{1 \rightarrow 2})}\big[(\dfrac{(1-P_{1 \rightarrow 0}-P_{1 \rightarrow 2})(P_{2 \rightarrow 1}+P_{2 \rightarrow 3})+(P_{1 \rightarrow 2}P_{2 \rightarrow 1})}{P_{2 \rightarrow 1}+P_{2 \rightarrow 3}})\rho_1^f\\&+(\dfrac{P_{1 \rightarrow 2}}{(P_{1 \rightarrow 0}+P_{1 \rightarrow 2})})\rho_2^f+\dfrac{P_{1 \rightarrow 2}P_{2 \rightarrow 3}}{P_{2 \rightarrow 1}+P_{2 \rightarrow 3}}{\rho_3^f}(t_3^f+1)\big]
   \label{fie3}
\end{split}
\end{align}
\end{widetext}

where $\rho^f_1$ is given in (\ref{rho1final}), $\rho^f_2$ in (\ref{rho2}), and $\rho^f_3$ in (\ref{mixed}).

To get $t_3^f$ we use a well-known technique described in \cite{traulsen3,mehdi}.
The time to fixation starting from $i$ mutant is given by the following formula,
\begin{equation}
	t^{f}_{i}=\sum_{j=1}^{N-1}\frac{\rho^f_j}{\rho^f_i}F_{ij}.
\label{fie4}
\end{equation}
where $F_{ij}$, given in (\ref{17}) in the Appendix (A), is the expected number of times,
the process which started from $i$ mutants, passes through
the state with $j$ mutants before reaching one of the absorbing states and $\rho^f_j$ is the fixation probability starting from $j$ mutants. 
Values of $F_{ij}$ depend on graphs (see the Appendix (A) for more details).

Fig. \ref{ft} shows that there is no difference in fixation times for various degrees of an initial mutant. The distribution of fixation times is shown in Fig. \ref{ft3}. 
We can see that the fixation times are distributed symmetrically around the average fixation time in both graphs. 
One notable point is that the average fixation time in the BA graph is higher than in the ER one. 

\subsection{Extinction Time}

\begin{figure*}
{\includegraphics[scale=0.29]{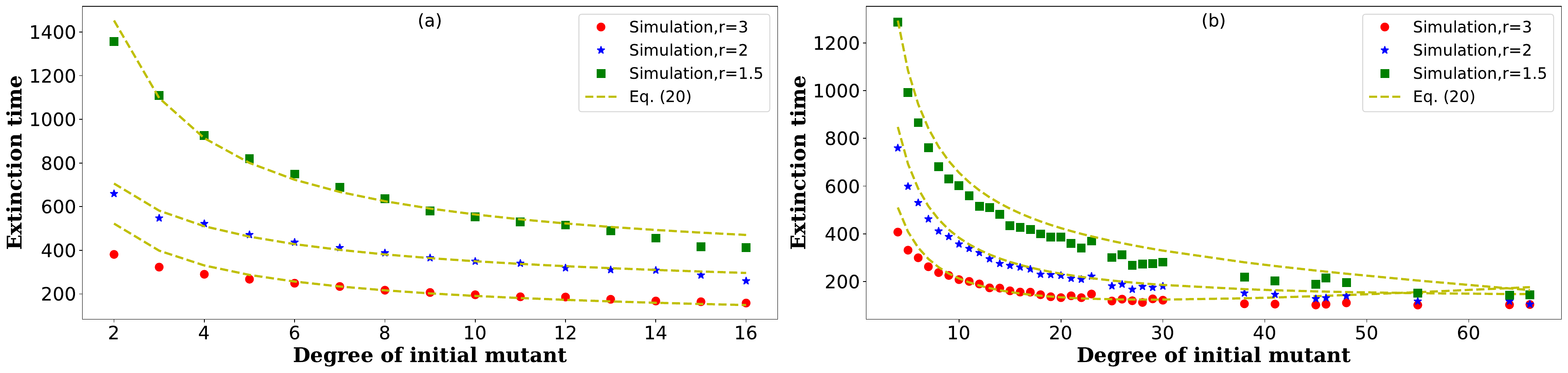}}
\centering
\caption{ Extinction time as a function of a degree of an initial mutant for three fitness values: 3, 2, 1.5 in (a) ER and (b) BA graphs 
with the size of 500 and the average degree of 8.}
\label{et1}
\end{figure*}

To find an analytical expression for the extinction time we use a similar approach as we did for the fixation time. Let $\rho_i^e=1-\rho_i^f$ 
be the extinction probability starting with $i$ mutants, and $P_{i \rightarrow j}$ the transition probability from $i$ to $j$ mutants. 
We apply the conditional expected value formula for $t_i^e$, the extinction time from a state with $i$ initial mutants, and get \cite{Antal} 

\begin{equation}
    \rho_1^e t_1^e=P_{1 \rightarrow 0}+P_{1 \rightarrow 1}\rho_1^e (t_1^e+1)+P_{1 \rightarrow 2}\rho_2^e (t_2^e+1) 
    \label{exe1}
\end{equation}
and
\begin{equation}
    \rho_2^e t_2^e=P_{2 \rightarrow 1}\rho_1^e (t_1^e+1)+P_{2 \rightarrow 2}\rho_2^e (t_2^e+1)+P_{2 \rightarrow 3}\rho_3^e (t_3^e+1)
    \label{exe2}
\end{equation}

These equations for the extinction time are similar to equations we derived for the fixation time. In (\ref{exe1}) there is an extra term $P_{1 \rightarrow 0}$ 
because if the number of mutants reaches zero, the extinction probability from this state is zero $t_0^e=0$ and the term $P_{1 \rightarrow 0}\rho_0^e (t_0^e+1)$ drops out, 
leaving only the $P_{1 \rightarrow 0}$ term. As before we find $\rho_2^e t_2^e$  from  (\ref{exe2}) and substitute it to (\ref{exe1}) and after some simplifications we get
\begin{widetext}
\begin{align}
\begin{split}
   t_1^e&=\dfrac{P_{2 \rightarrow 1}+P_{2 \rightarrow 3}}{\rho_1^e((P_{1 \rightarrow 0}+P_{1 \rightarrow 2})(P_{2 \rightarrow 1}+P_{2 \rightarrow 3})-P_{2 \rightarrow 1}P_{1 \rightarrow 2})}\big[P_{1 \rightarrow 0}+(\dfrac{(1-P_{1 \rightarrow 0}-P_{1 \rightarrow 2})(P_{2 \rightarrow 1}+P_{2 \rightarrow 3})+(P_{1 \rightarrow 2}P_{2 \rightarrow 1})}{P_{2 \rightarrow 1}+P_{2 \rightarrow 3}})\rho_1^e\\&+(\dfrac{P_{1 \rightarrow 2}}{(P_{1 \rightarrow 0}+P_{1 \rightarrow 2})})\rho_2^e+\dfrac{P_{1 \rightarrow 2}P_{2 \rightarrow 3}}{P_{2 \rightarrow 1}+P_{2 \rightarrow 3}}{\rho_3^e}(t_3^e+1)\big]
   \label{exe3}
\end{split}
\end{align}
\end{widetext}

Here again the idea is that we consider local information and an effect of the initial degree in states $1,2,3$ 
and then we disregard the information about the  initial degree. 
The time to extinction, starting from $i=3$ mutants is given by the following expression \cite{traulsen3,mehdi},
\begin{equation}
	t^{e}_{3}=\sum_{j=1}^{N-1}\frac{\rho^e_j}{\rho^e_3}F_{3j}.
\label{112}
\end{equation}
\begin{figure}
{\includegraphics[scale=0.50]{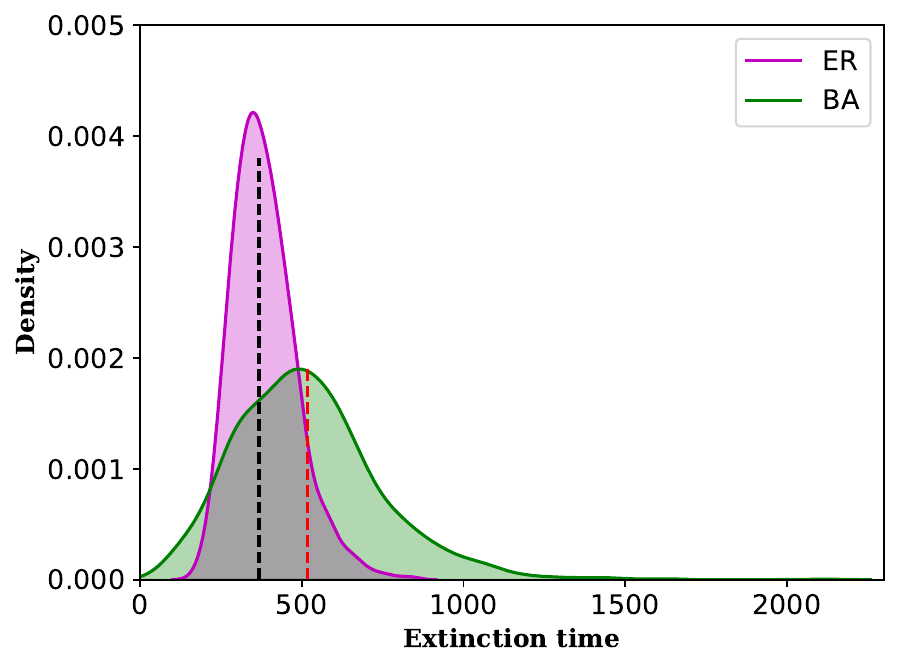}}
\centering
\caption{Distribution of extinction times of a mutant with respect to initial placements on ER and BA graphs with the size of $500$ 
and the average degree of $8$. The fitness is $2$. The dash lines show the average extinction times.}
\label{et2}
\end{figure}
\begin{figure}
{\includegraphics[scale=0.5]{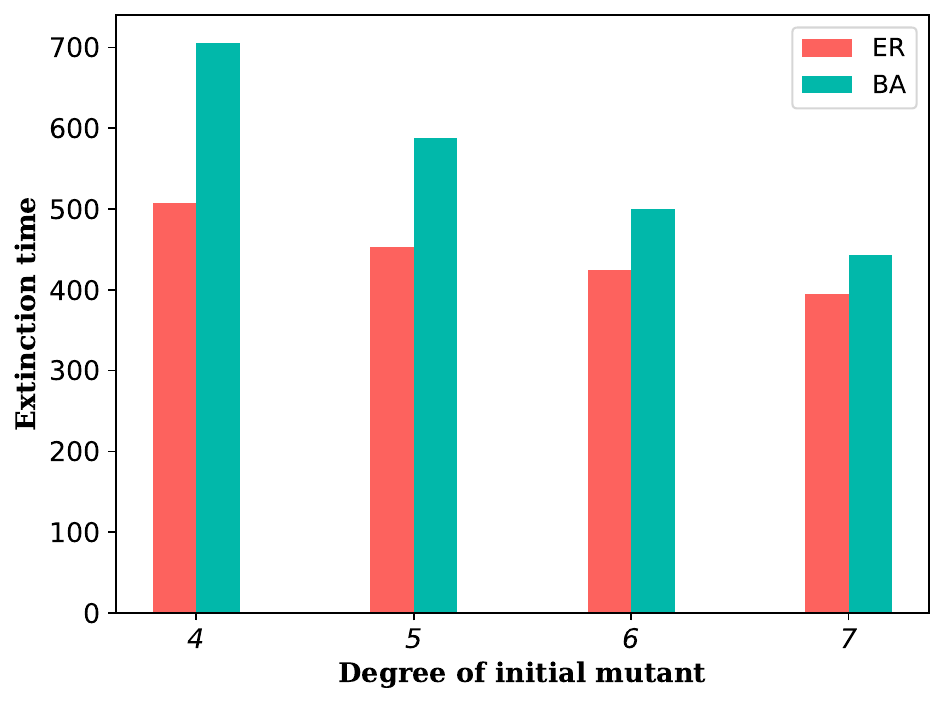}}
\centering
\caption{Extinction time for various initial mutants with the same degree in both networks. The size of the networks is $500$, the average degree $8$ and $r=2$.}
\label{ets}
\end{figure} 
 \begin{figure*}
{\includegraphics[scale=0.26]{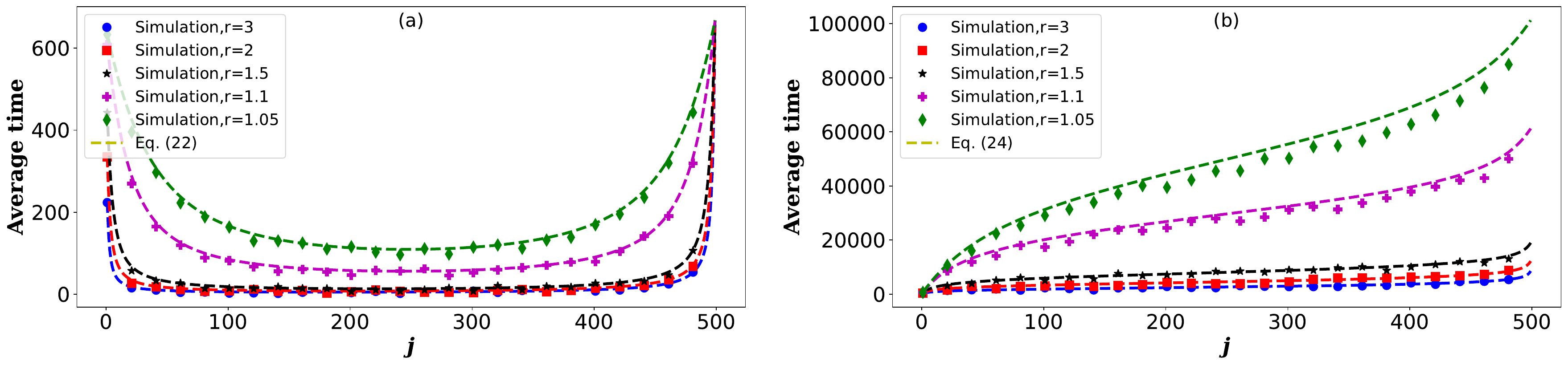}}
\centering
\caption{ a) The expected number of times, the population started with one mutant passes through the state with $j$ mutants before fixation, $t^f_{1,j}$, eq. (22) 
and b) the expected time until the number of mutants reaches $j$ from $1$, $T^f_{j}$, eq. (17), for various values of $r$ in the ER graph with $N=500$ 
and average degree of $8$.}
\label{fixtimee}
\end{figure*}

\begin{figure*}
{\includegraphics[scale=0.26]{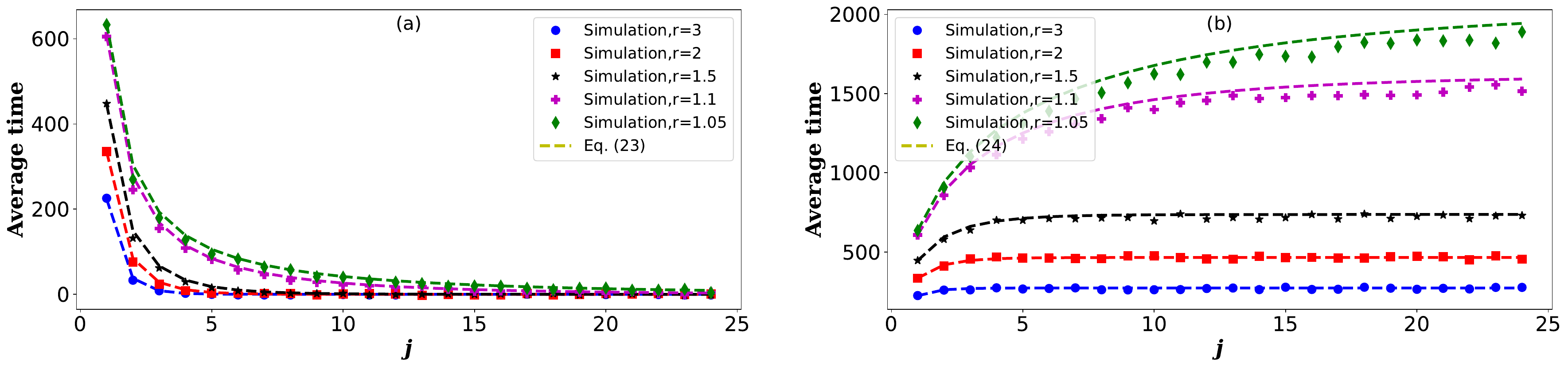}}
\caption{ a) The expected number of times, the population started with one mutant passes through the state with $j$ mutants before extinction, $t^e_{1,j}$, eq. (23) 
and b) the expected time until the number of mutants reaches $j$ from $1$, $T^e_{j}$, eq. (17), for various values of $r$ in the ER graph with $N=500$ 
and average degree of $8$.}
\label{extimee}
\end{figure*}

We present results of stochastic simulations and our analytical approach in Fig. \ref{et1}. 
It appears that as the initial mutant's degree increases, extinction time decreases. Essentially, 
as the initial mutant degree increases, it becomes more likely to select one of its neighbors to replace it. 
As a result, mutants with higher degrees are more likely to be wiped out in a short period of time. In the BA graph, for two initial mutants 
with a large difference in degrees, we have completely different extinction times due to the average degree of the network.  
It is worth noting that for vertices with high degrees, the extinction 
time is almost the same for all values of r. In other words, if the process starts from a hub, regardless of $r$, it will soon become extinct.
In Fig. \ref{et2}, we can see the distribution of extinction times for both graphs. In the ER graph, 
extinction times are distributed around an average time, while in the BA one, 
the distribution is more extended. ER graphs exhibit lower average extinction times than BA ones.

In Fig. \ref{ets}, the extinction time for various initial mutants with equal degrees is presented. 
Observe that for initial mutants with the same degree, the one in the BA graph exhibits a longer extinction time. 

\section{Fixation and extinction times - discussion}
\label{dis}

Here we examine in more detail the evolution of a population starting with one mutant.  
Fig. \ref{fixtimee} and Fig. \ref{extimee} show the path to extinction and fixation of a mutant in the ER graph with a size of $500$ and an average degree of 8. 
We derive approximate formulas for the expected number of times the process which started from $i$ mutants, 
passes through the state with $j$ mutants before reaching fixation and extinction respectively (details of derivation are given in the Appendix (A)),

\begin{equation}
t^f_{i,j}=
\begin{cases}
\dfrac{{(r^j-1)}^2 (r^{N-i}-1)r^{i-j}}{(r-1)(r^N-1)(r^i-1)\frac{j}{N-i+rj}\frac{(N-j)(p-\frac{2}{N})}{Np}} \hspace{5pt} j \leq i \\[15pt]

\dfrac{(r^{N-j}-1)(r^j-1)}{(r-1)(r^N-1)\frac{j}{N-i+rj}\frac{(N-j)(p-\frac{2}{N})}{Np}}\hspace{39pt} j>i
\end{cases}
\end{equation}

\begin{equation}
t^e_{i,j}=
\begin{cases}
\dfrac{{(r^j-1)} (r^{N-i}-1)(r^N-r^j)r^{i-j}}{(r-1)(r^N-1)(r^N-r^i)\frac{j}{N-i+rj}\frac{(N-j)(p-\frac{2}{N})}{Np}} \hspace{5pt} j \leq i \\[15pt]

\dfrac{(r^N-r^j)(r^{N-j}-1)(r^i-1)}{(r-1)(r^N-1)(r^N-r^i)\frac{j}{N-i+rj}\frac{(N-j)(p-\frac{2}{N})}{Np}} \hspace{5pt} j>i
\end{cases}
\end{equation}

Denote by $T_j^e$ and $T_j^f$ the expected time it takes a single mutant to take over $j$ nodes in a network before extinction and fixation respectively. 
For a single mutant to be fixed, the Moran process must pass through the state with $j$ mutants.  
Thus, $T_j^f$ is the time needed for a single mutant to be fixed minus the time needed for $j$ mutants to be fixed \cite{mehdi2}. Therefore we have,
\begin{equation}
\begin{split}
  T^f_{j}=t^{f}_1-t^{f}_j=\sum_{l=1}^{N-1}t^f_{1,l}-\sum_{l=1}^{N-1}t^f_{j,l},\\ T^e_{j}=t^{e}_1-t^{e}_j=\sum_{l=1}^{N-1}t^e_{1,l}-\sum_{l=1}^{N-1}t^e_{j,l}.
\end{split}
\end{equation}

We have performed stochastic simulations to see how accurate these analytical formulas are. For $t_1^f$ and $t_1^e$ 
we use our analytical approximation in (\ref{fie3}) and (\ref{exe3}). Panels (a) in Fig. \ref{fixtimee} and Fig. \ref{extimee} 
show the expected number of times, the population started with one mutant passes through the state with $j$ mutants
before fixation and extinction respectively while panels (b) show the value of $T_j^f$ and $T_j^e$. 
The results are for the ER network with a size of $500$ and an average degree of $8$. 
We present here only results for the ER network, results for the BA network are the same. 
We observe that the process before extinction spends most of its time in states with few mutants. 
This means that during the process of the extinction of mutants, a few vertices get mutated 
and if the number of mutants increases, there is a higher chance of them to be fixed. 
Therefore, the initial placement of mutants plays an important role in the extinction of mutants. 
For fixation, unlike extinction, the beginning and the end of the process can have a significant impact on the fixation time.
At the beginning of the process when few mutants are available, the probability of choosing them to reproduce is low, so it takes a long time for mutants 
to increase their number. Furthermore, with fewer residents in the population, it takes longer time to replace them with one of the mutants' offspring 
at the end of the process. Hence it doesn't matter where the mutant is initially placed because the start and end of the process are equally important 
for the fixation time.

We observe that our analytical expressions are less accurate for smaller values of $r$ and are becoming better as $r$ increases. 
We assumed that the initial mutant degree influences only the probability of transition between one to two and two to three mutants. 
However, there is a possibility that the number of mutants increases and then decreases again to one. 
Hence, the mutant may be different from the starting mutant, leading to an error in our approximation. 
For large $r$, there is a low probability that the number of mutants will decrease to one. Therefore, when there are more mutants,
they are more likely to increase rather than decrease their number. Consequently, our approximation is better for larger $r$.     

\section{Discussion}
\label{summ}

To summarize, we examined the impact of the degree of an initial mutant on the fixation probability and fixation and extinction times 
in Moran processes in structured populations. We performed computer simulations and developed a general mean-field approximation 
approach on graphs which enabled us to get analytical formulas. In this way we extended to arbitrary fitness values results of \cite{Antal2,Sood,maciejewski}, 
where the authors provided good aproximations for cases with mutant fitness close to that of a resident. 

We showed that the fixation probability depends on the vertex degree at which a mutant is introduced. 
Increasing the degree of the initial mutant makes it more likely for the mutant to be replaced by one of its neighbors. 
There was no correlation between the degree of the initial mutant and the time it took for the mutant to take over the entire population. 
The fixation process depends on the entire population, not on the first mutant.
We found that, unlike fixation time, extinction time significantly depends on the degree of the first mutant - a mutant with fewer connections needs more time 
to become extinct. Furthermore, we observed that the extinction time for an initial mutant with a small degree is significantly influenced by its fitness, 
but as the degree of the first mutant increases, extinction times become independent on the fitness value.

It is important to study the impact of the initial mutant's degree on the fixation probability and fixation and extinction times
in frequency-dependent Moran processes of spatial evolutionary games, where fitness is derived from game competitions.  

{\bf Acknowledgments}: This project has received funding from the European Union’s Horizon 2020 research and innovation program 
under the Marie Sk\l odowska-Curie grant agreement No 955708. Computer simulations were made with the support 
of the Interdisciplinary Center for Mathematical and Computational Modeling of the University of Warsaw (ICM UW).

\section*{Appendix A}
\label{AP}
\label{evo}
We consider here a reduced Markov chain for the Moran process on the graph with $N+1$ states representing the number $i$ of mutants in the population. 
The transition matrix can be then written as follows:
\begin{equation}
P_{ij}=p_{i \rightarrow i+1} \delta_{i+1,j}+p_{i \rightarrow i-1} \delta_{i-1,j} + q_i \delta_{i,j}\;,
\label{12}
\end{equation}
where $q_{i}=1-p_{i \rightarrow i+1}-p_{i \rightarrow i-1}$. 

Our Markov chain has two absorbing states: $i=0$ and $i=N$. To calculate absorbing probabilities and expected absorbing times we use 
the general method described in \cite{grinstead}. We rewrite the transition matrix as follows:
\begin{equation}
P= 
\begin{pmatrix}
  Q & R\\ 
  0 & I
\end{pmatrix},
\label{13}
\end{equation} 

where $Q$ contains transition probabilities between transient states and $R$ contains transition probabilities from transient states to absorbing ones, 
and $I$ is the identity matrix. $F=(I-Q)^{-1}$ is called the fundamental matrix. It can be shown that $F_{ij}$ is the expected number of times, 
the process which started from $i$ mutants, passes through the state with $j$ mutants before reaching one of the absorbing states.  
Then it follows that expected absorption times are given by summations,
\begin{equation}
t_i=\sum_{j=1}^{N-1}F_{ij}.
\label{14}
\end{equation} 

Expected number of times the process which started from $i$ mutants, passes through the state with $j$ mutants before reaching 
fixation and extinction respectively can be written as follows \cite{traulsen3},
\begin{equation}
\begin{array}{rl}
t^{f}_{i, j}=\frac{\rho^f_j}{\rho^f_i}F_{ij},\\
t^{e}_{i, j}=\frac{\rho^e_j}{\rho^e_i}F_{ij}.
\end{array}
\label{16}
\end{equation}

Hence expected fixation and extinction times are given by the following expressions,
\begin{equation}
\begin{array}{rl}
t^{f}_{i}=\sum_{j=1}^{N-1}\frac{\rho^f_j}{\rho^f_i}F_{ij},\\
	t^{e}_{i}=\sum_{j=1}^{N-1}\frac{\rho^e_j}{\rho^e_i}F_{ij}.
\end{array}
\label{17}
\end{equation}

Now we turn our attention to graphs. It can be shown that if the Moran process on a graph satisfies the condition 
$p_{i \rightarrow i+1} = rp_{i \rightarrow i-1}$, then we have \cite{mehdi},

\begin{equation}
  F_{ij}=
  \begin{cases}
    \frac{(r^{j}-1)(r^{N-i}-1)}{p_{j \rightarrow j-1}(r-1)(r^N-1)}, j\leq i \\[15pt]
     \frac{r^{j-i}(r^{i}-1)(r^{N-j}-1)}{p_{j \rightarrow j-1}(r-1)(r^N-1)}, i<j
  \end{cases}
\label{17}
\end{equation}

The authors of \cite{mehdi} used mean-field techniques and argue that $p_{i \rightarrow i+1} = rp_{i \rightarrow i-1}$ is a good approximation for
Erd\"{o}s-R\'{e}nyi and Barab\'{a}si-Albert graphs. We present here their arguments and results.

\begin{itemize}
\item Erd\"{o}s-R\'{e}nyi graph (ER)

In the ER graph with $N$ vertices and $j$ mutants, the probability of a mutant being selected for the reproduction is given by $\frac{rj}{rj+N-j}$. 
Although it may seem that this mutant would have $(N-j)p$ residents in its neighborhood, the population of mutants tends to form clusters, 
making it likely for a selected mutant to have at least two mutant neighbors. Thus, the probability of a selected mutant being connected to a resident 
is approximately $(p-\frac{2}{N})$, and the average number of residents connected to a selected mutant is ($N-j)(p-\frac{2}{N})$. Then

\begin{equation}
p_{j \rightarrow j+1} =rp_{j \rightarrow j-1}=\frac{rj}{N-i+rj}\frac{(N-j)(p-\frac{2}{N})}{(N-1)p}.
\label{18}
\end{equation}

\item  Barab\'{a}si-Albert graph (BA)

Here again, the probability for a mutant to be chosen for the reproduction is $\dfrac{rj}{rj+N-j}$. For a state with $j$ mutants, $I_j$ represents 
the average number of edges that connect different species, also known as interface edges. On average, each mutant (resident) has $<k>$ connections, 
and $\frac{I_j}{j}$  ($\frac{I_j}{(N-j)}$) interface edges. As a result, the probability for a mutant (resident) offspring to replace 
one of its resident (mutant) neighbors is $\frac{\frac{I_j}{j}}{m}$  ($\frac{\frac{I_j}{(N-j)}}{m}$). It follows that

\begin{equation}
p_{j \rightarrow j+1} =rp_{j \rightarrow j-1}=\frac{rI_j}{<k>(N-j+rj)}.
\label{20}
\end{equation}

In order to find $I_j$ when there are $i$ mutants, it is evident that every resident node has at least $m$ edges and an average 
of $\frac{I_j}{N-j}$ interface edges. 
If one resident is changed to a mutant, the number of mutants increases by 1. 
Therefore, we can obtain a recursive relationship for the number of interface edges as follows:

\begin{equation}
I_{j+1}=I_j+m-\dfrac{I_j}{N-j}.
\label{21}
\end{equation}

Assuming $I_0=0$ we have:
\begin{equation}
I_j=m\sum_{i=1}^{j}\frac{N-j}{N-i}.
\label{22}
\end{equation}

\end{itemize}

\section*{Appendix B}

The effect of the degree of an initial mutant on the fixation probability has been previously studied by Antal, Redner, and Sood in \cite{Antal2}. 
They examined how initial placement impacts fixation probability in various dynamics. 
In particular, for the birth-death Moran process (biased invasion process) with the mutant fitness equal to $1$ and the resident fitness set to $1-s$,
the authors derived the following approximate formula for the fixation probability:

\begin{equation}
    \rho_1^f(\omega_{-1})=\dfrac{1-e^{-sN\omega_{-1}/(1-s/2)}}{1-e^{-sN/(1-s/2)}}
    \label{eq35}
\end{equation}

where $\omega_{-1}=\frac{1}{k\mu_{-1}N}$ and $\mu_{-1}=\sum_{k}^{}\frac{1}{k}\rho(k)$ is the average inverse degree ($\rho(k)$ is the degree distribution) \cite{Antal2}.

\begin{figure}[H]
    \centering
    \includegraphics[scale=0.4]{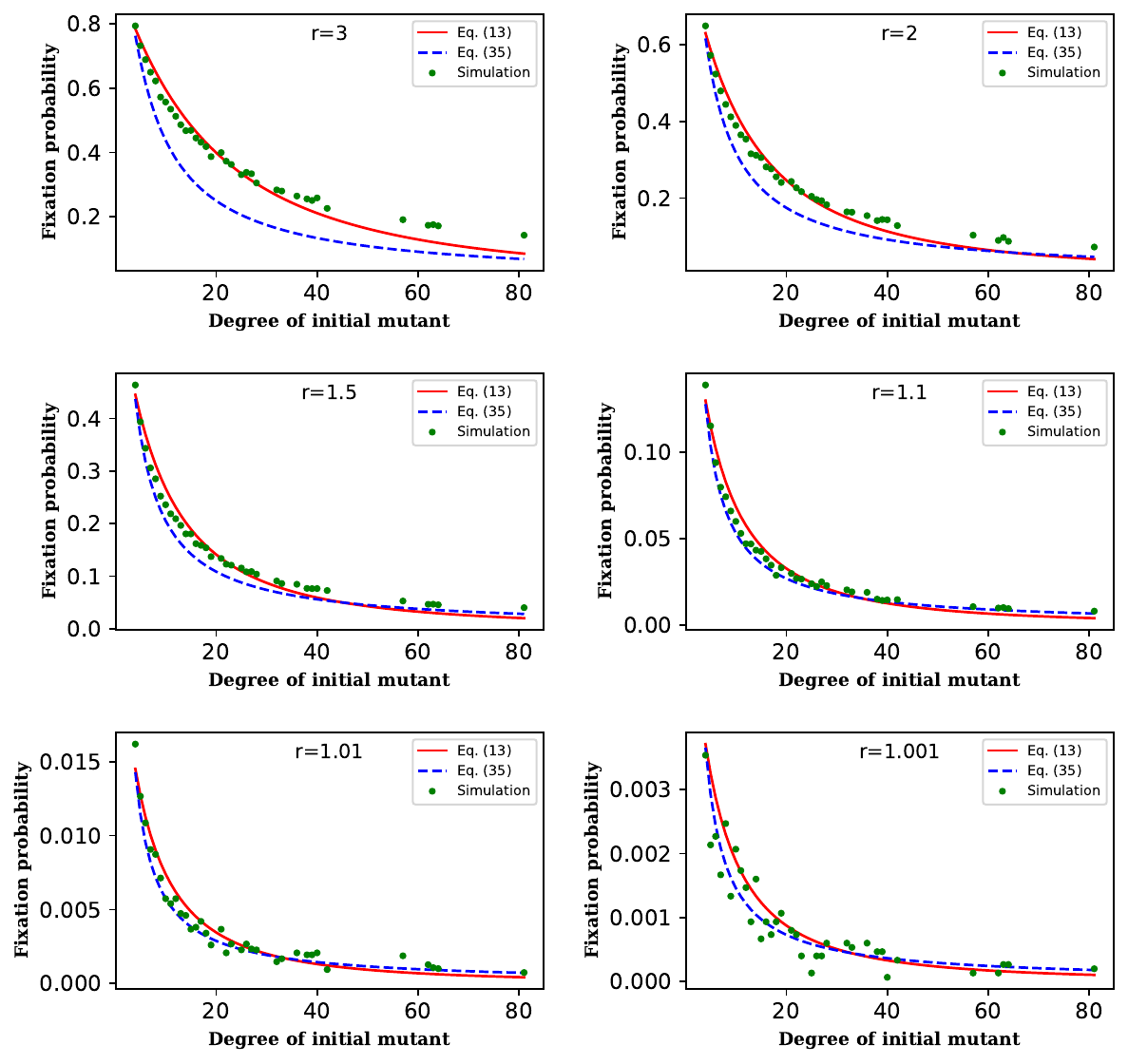}
    \caption{Fixation probability as a function of the degree of an initial mutant for various fitness values. The red line is our approximation given by (\ref{rho1final}),
 the blue dashed line is an approximation given by (\ref{eq35}), and the green dots show the result of simulations. In equation (\ref{eq35}) $s=1-1/r$. }
    \label{fig:comp1}
\end{figure}

Fig. \ref{fig:comp1} compares the results of our approximation given in eq. (\ref{rho1final}) with eq. (\ref{eq35}) for various fitness values. 
We calculated $\mu_{-1}$ numerically.
To make the connection between the fitness $r=1+s>1$ for mutants and fitness $1$ for residents in our paper with \cite{Antal2}, 
where fitness of mutants is $1$ and that of residents $1-s$, we divide the fitnesses in (\ref{eq35}) by $1-s$ (fixation probability is invariant under such a scaling).
Therefore in (\ref{eq35}), the fitness of mutants is $r=\frac{1}{1-s}>1$ for $0<s<1$ and that of resident is $1$.
In this way the value of $s$ in (\ref{eq35}) is $1-\frac{1}{r}$.  

We see that our approximation performs well for all values of $s>0$, while eq. (\ref{eq35}) is more accurate in cases where $r$ is close to $1$, $(s<<1)$.

\begin{figure}[H]
    \centering
    \includegraphics[scale=0.4]{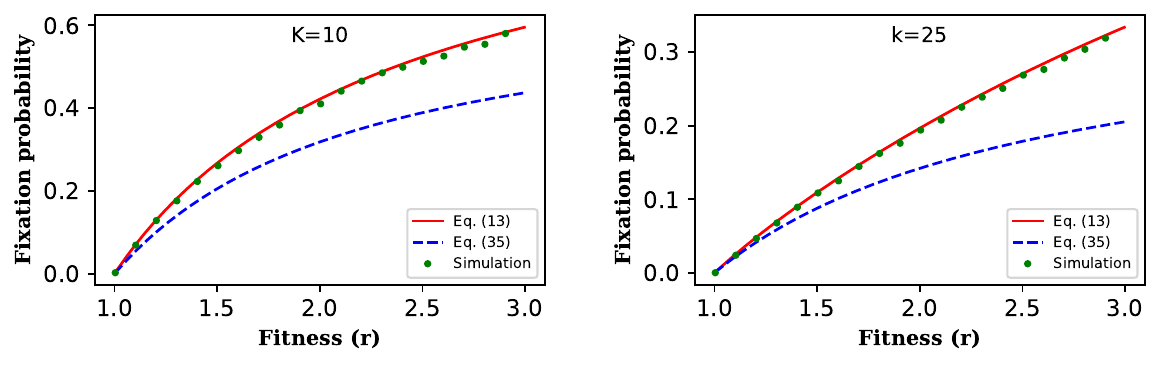}
    \caption{The fitness effect on fixation probability for an initial mutant with degrees of 10 and 25. The red line is our approximation given by (\ref{rho1final}), the blue dashed line is an approximation given by (\ref{eq35}), and the green dots show the result of simulations. In equation (\ref{eq35}) $s=1-1/r$.}
    \label{fig:comp2}
\end{figure}

For a more direct comparison, we fixed the degree of an initial mutant and varied the fitness $r$ from $1$ to $3$. 
Fig. \ref{fig:comp2} shows the results for initial mutants with degrees of $10$ and $25$. For $r$ close to $1$, both approximations perform well. 
However, as $r$ increases, our approximation in equation (\ref{rho1final}) outperforms equation (\ref{eq35}).

\end{document}